# Nanoparticle Deposition Techniques for Silica Nanoparticles: Synthesis, Electrophoretic Deposition, and Optimization- A review

Author List
a) Srabani Karmakar, Department of Materials Science & Engineering, University of Utah.
b) Milind Deo, milind.deo@utah.edu, Departments of Chemical Engineering, Professor, University of Utah
c) Imteaz Rahaman, Department of Electrical & Computer Engineering, University of Utah
d) Swomitra Kumar Mohanty swomitra@chemeng.utah.edu Associate Professor, Departments of Chemical Engineering, Department of Materials Science Engineering, University of Utah

# Highlights

- **Stöber Method for Enhanced Nanoparticle Synthesis:** Highlights the Stöber method as a leading technique for producing monodisperse silica nanoparticles with controlled size, morphology, and surface properties, ideal for advanced material applications.

- **Electrophoretic Deposition (EPD) as a Game-Changer:** EPD is presented as a scalable, cost-effective, and versatile deposition method for non-metallic nanoparticles like silica nanoparticles ($SiO_2$), offering room-temperature processing, uniform deposition, and compatibility with complex geometries.

- **Comparative Advantages of EPD of silica nanoparticles ($SiO_2$) Over Conventional Methods:** A detailed comparison demonstrates the advantages of EPD over techniques like CVD, ALD, and spin coating in terms of scalability, uniformity, and material compatibility, particularly for non-metallic materials.

- **AI-Driven Optimization for Enhanced EPD of $SiO_2$:** The review explores how the integration of Artificial Intelligence (AI) and active learning into EPD will enable predictive optimization, real-time monitoring, and dynamic adjustments, which can significantly improve film quality and deposition efficiency.

## Abstract

Silica nanoparticles have emerged as key building blocks for advanced applications in electronics, catalysis, energy storage, biomedicine, and environmental science. In this review, we focus on recent developments in both the synthesis and deposition of these nanoparticles, emphasizing the widely used Stöber method and the versatile technique of electrophoretic deposition (EPD). The Stöber method is celebrated for its simplicity and reliability, offering precise control over particle size, morphology, and surface properties to produce uniform, monodisperse silica nanoparticles that meet high-quality standards for advanced applications. EPD, on the other hand, is a cost-effective, room-temperature process that enables uniform coatings on substrates with complex geometries. When compared to traditional techniques such as chemical vapor deposition, atomic layer deposition, and spin coating, EPD stands out due to its scalability, enhanced material compatibility, and ease of processing. Moreover, Future research should integrate AI-driven optimization with active learning to enhance electrophoretic deposition (EPD) of silica nanoparticles, leveraging predictive modeling and real-time adjustments for improved film quality and process efficiency. This approach promises to accelerate material discovery and enable scalable nanofabrication of advanced functional films.

## 1. Introduction

Nanoparticle deposition techniques are fundamental to the fabrication of nanoscale architectures and functional devices, playing a pivotal role in modern technological advancements [1], [2]. These methods enable precise control over the spatial arrangement of nanoparticles on a variety of substrates, which is essential for tailoring optical, electronic, magnetic, and catalytic properties in materials [2]. Such precise control has led to innovations across diverse fields including electronics, catalysis, energy storage, biosensors, and healthcare [3], [4], [5]. For instance, high-performance sensors, memory devices, biomedical implants, and drug delivery systems all rely on the meticulous design of nanoparticle assemblies [6], [7]. As the demand grows for materials that combine high precision with scalability and efficiency, enhancing deposition methods becomes increasingly critical. Recently, transformative tools such as artificial intelligence (AI) and active learning have been introduced to optimize these processes, promising breakthroughs in predictive modeling, real-time monitoring, and dynamic adjustment of deposition parameters [8], [9].

Over the past two decades, an array of nanoparticle deposition methods has been developed, each with unique advantages regarding resolution, material compatibility, and scalability. Traditional techniques such as physical vapor deposition (PVD), chemical vapor deposition (CVD), and various electrochemical methods have been instrumental in semiconductor manufacturing and surface coatings [10], [11], [12], [13]. However, these methods often face limitations when applied to non-metallic materials like silicon dioxide ($SiO_2$), particularly in terms of achieving uniform large-area films and controlled thicknesses. In contrast, more advanced methods including atomic layer deposition (ALD) and electron beam lithography (EBL) offer superior precision but at the expense of high cost and lengthy processing times [14], [15]. Scalable alternatives like dip coating, spin coating, inkjet printing, and self-assembly, while cost-effective, may not always provide the level of uniformity required for advanced applications [16]. On the other hand, Electrophoretic deposition (EPD) offers significant advantages over conventional methods like dip coating, spin coating, and inkjet printing, providing superior uniformity, higher packing density, and precise control over film thickness [17]. The effectiveness of EPD relies heavily on the colloidal stability and surface charge of the deposited nanoparticles, making silica ($SiO_2$) an ideal candidate due to its tunable surface chemistry and consistent negative charge in suspension [17], [18]. These $SiO_2$ nanoparticles are most efficiently synthesized via the Stöber method, which ensures monodisperse, size-controlled particles with excellent reproducibility [19]. By combining the precision of EPD with the high-quality $SiO_2$ nanoparticles produced by the Stöber process, researchers can achieve highly uniform and scalable thin-film nanostructures for cutting-edge applications.

This review synthesizes recent developments in both the synthesis and deposition of $SiO_2$ nanoparticles, focusing on the strengths and limitations of current approaches while exploring innovative solutions for future advancements. We begin by examining the Stöber method as an efficient route to high-quality, uniform $SiO_2$ nanoparticles, followed by an in-depth discussion of electrophoretic deposition (EPD) as a versatile, room-temperature technique for depositing non-metallic nanoparticles on complex substrates. The review further evaluates the integration of AI and active learning into the EPD process, highlighting how these techniques optimize deposition parameters through predictive modeling and real-time adjustments. This work aims to offer valuable insights and a comprehensive framework to guide future research and industrial applications in nanofabrication and functional material design by providing a comparative analysis of various deposition methods and emphasizing the transformative potential of AI-driven optimization. Fig 1 shows the complete flowchart of this review article.

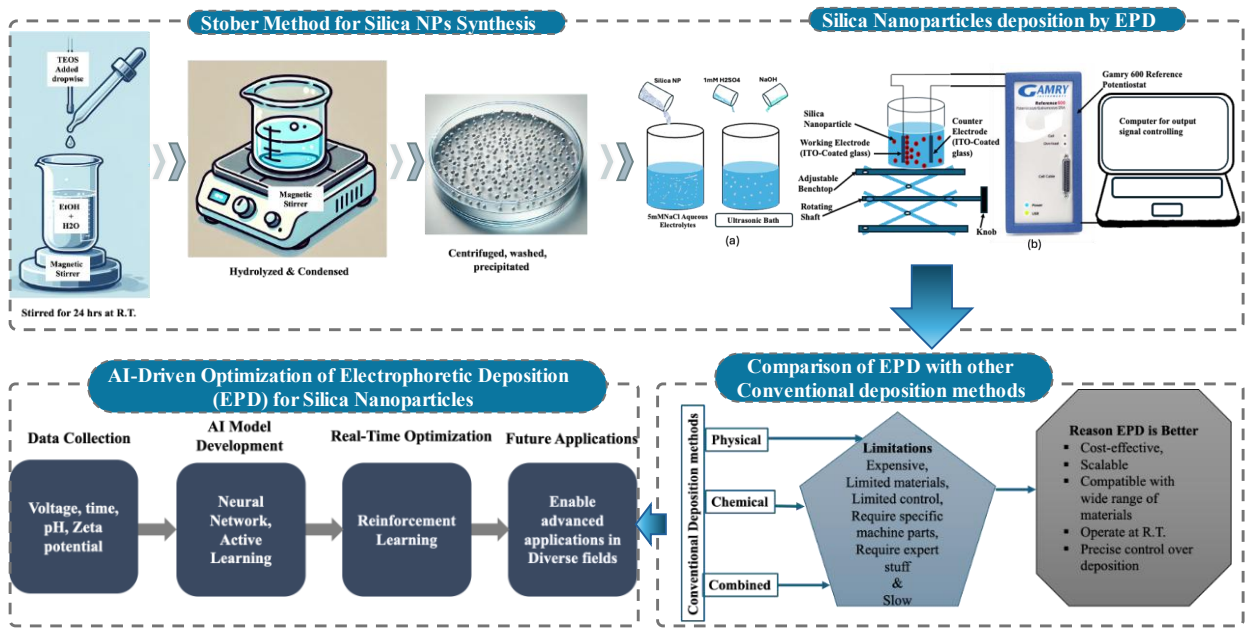

Fig1: Integrated Workflow for Silica Nanoparticle Synthesis, Deposition by Electrophoretic deposition (EPD) method, and AI-driven optimization: From Synthesis to Advanced Applications.

## 1.1. Importance of Nanoparticles for Fabrication of Nanostructured Systems

As nanotechnology emerged as one of the most transformative fields in modern nanoparticle deposition methods, the synthesis and application of nanoparticles—ranging from 1 to 100 nm in diameter—have found widespread use in environmental science, agriculture, biotechnology, and biomedicine [20]. A wide range of applications employ nanoparticles, such as antibacterial agents [21], functional food additives [22], wastewater treatment [23], and remediation of environmental pollutants [24], as well as fabrication materials for nanostructured systems such as nanofluidic devices [25]. Fig 2 shows different percentages of nanomaterial application in different fields. Figure 2 shows how the demand for nanomaterials applications in materials science and engineering departments grows significantly. Their unique properties, such as biocompatibility, antibacterial activity, and efficient drug delivery, have required nanoparticles in biotechnological applications [26]. This growing demand for nanoparticle applications across multiple disciplines emphasizes the importance of nanoparticle patterning methods.

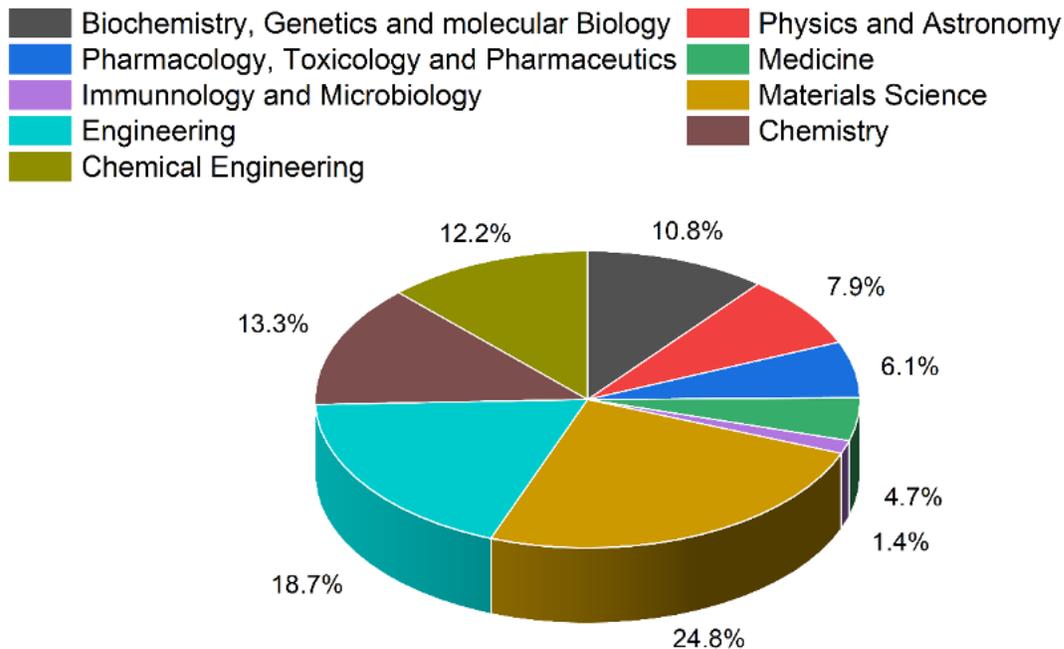

Fig 2: Nanomaterials for materials science and other engineering and materials science applications (adapted from [27].)

The physical characteristics of nanoparticles are profoundly different from those of bulk materials due to their small size and large surface area. These differences arise because, as particle dimensions approach the nanoscale, quantum effects and surface phenomena become dominant. For instance, size-dependent quantum confinement affects the electronic and optical behavior of nanoparticles, leading to enhanced catalytic activity, tunable optical absorption, and modified electrical conductivity. Furthermore, the disruption of periodic boundary conditions in the crystalline structure—particularly when the size of nanoparticles approaches the de Broglie wavelength or the wavelength of light—results in distinct plasmonic, electronic, and catalytic properties [28]. These properties make nanoparticles indispensable for applications in medicine, energy storage, and nanoelectronics [29].

## 2. Silica Nanoparticle Synthesis Processes

Silica nanoparticles ($SiO_2$) have emerged as one of the most versatile and widely used nanomaterials, owing to their unique properties such as high surface area, tunable porosity, chemical stability, and biocompatibility [30]. These characteristics make them indispensable in a wide range of applications, including drug delivery, catalysis, coatings, biosensors, and energy storage [31]. The synthesis of silica nanoparticles is a critical step that determines their size, morphology, surface chemistry, and functionality [32], [33]. Over the years, various synthesis methods have been developed, each offering distinct advantages and limitations [34], [35]. This section provides a comprehensive overview of the most prominent silica nanoparticle synthesis processes, including the Stöber method, microemulsion, flame spray pyrolysis, biological synthesis, and laser ablation. Fig 3 shows different methods of synthesis of silica nanoparticles. Each method is discussed in detail, highlighting its mechanisms, key parameters, advantages, limitations, and applications.

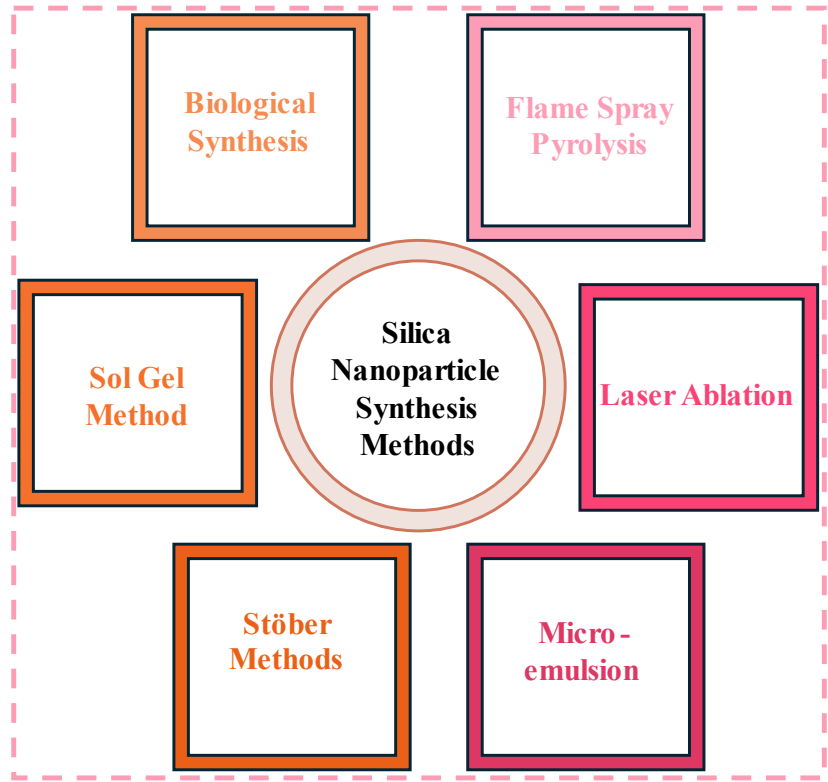

Fig 3: Conventional methods of synthesis of Silica Nanoparticles adapted from [36].

## 2.1. Microemulsion Method

The microemulsion method involves the formation of nanoscale water-in-oil droplets, which act as microreactors for the synthesis of silica nanoparticles [37]. In this method, water, oil, and a surfactant are mixed to form stable nanodroplets. TEOS is then added to the microemulsion, where it undergoes hydrolysis and condensation within the confined environment of the nanodroplets. The size of the nanoparticles is controlled by the size of the nanodroplets, which can be tuned by adjusting the surfactant concentration and the water-to-oil ratio [38]. The microemulsion method produces nanoparticles with a narrow size distribution and allows for the synthesis of complex nanostructures, such as core-shell particles [39]. However, the method requires the use of surfactants, which can be costly and may limit its scalability [38].

## 2.2 Flame Spray Pyrolysis

Flame spray pyrolysis is a high-temperature synthesis method that involves the combustion of a precursor solution to produce silica nanoparticles [40]. In this method, a precursor solution (e.g., TEOS in ethanol) is atomized into fine droplets, which are then ignited in a flame. The high temperature of the flame causes the precursor to decompose and react, leading to the rapid formation of silica nanoparticles. The nanoparticles are collected on a substrate or in a filter [40], [41]. Flame spray pyrolysis is a highly scalable method that can produce large quantities of silica nanoparticles in a short amount of time [41]. However, the method requires high energy consumption and offers limited control over the morphology of the nanoparticles [42].

## 2.3 Biological Synthesis

Biological synthesis involves the use of microorganisms or plant extracts to produce silica nanoparticles through bio-mediated processes [43], [44]. This method leverages the natural ability of certain organisms, such as diatoms, to catalyze the formation of silica nanoparticles from silicon precursors. The process begins with the bio-silicification of silicon precursors by microorganisms or plant extracts, followed by the extraction and purification of the silica nanoparticles [45]. Biological synthesis is an eco-friendly and sustainable approach that produces biocompatible nanoparticles, making it ideal for biomedical applications such as drug delivery and imaging [46]. However, the method typically yields low quantities of nanoparticles and offers limited control over particle size and morphology [47].

## 2.4 Laser Ablation

Laser ablation is a relatively recent and innovative technique for synthesizing silica nanoparticles. In this method, a high-energy laser beam is focused on a silicon target submerged in a liquid medium (e.g., water or ethanol) [48]. The laser energy causes the silicon target to vaporize, and the resulting silicon vapor condenses to form silica nanoparticles in the liquid medium. The size and morphology of the nanoparticles can be controlled by adjusting parameters such as laser power, pulse duration, and the properties of the liquid medium[48], [49]. Laser ablation offers several advantages, including the ability to produce highly pure nanoparticles without the need for chemical precursors or surfactants. The method also allows for the synthesis of nanoparticles with unique morphologies and surface properties [50]. However, laser ablation is a relatively expensive technique and may not be suitable for large-scale production [51]. Table 1 shows all the synthesis processes in a brief discussion.

Table 1: Key Synthesis Methods for Silica Nanoparticles [37], [40], [43], [44], [48].

| Synthesis Method | Description | Advantages | Limitations |
|---|---|---|---|
| Biological Synthesis | Uses plant extracts, microorganisms, or biomolecules for eco-friendly synthesis. | Sustainable; cost-effective; mild conditions. | Low yield, Slow reaction time, Poor size control, Requires bio-templates (e.g., diatoms) |
| Flame Spray Pyrolysis | Combustion of silicon precursors (e.g., $SiCl_4$ or TEOS) in a flame. | Scalable; high-purity NPs; rapid synthesis. | High energy cost, Agglomeration issues, Requires high temps (>1000°C), Limited surface functionalization |
| Laser Ablation | Laser pulses vaporize a silicon target in a liquid or gas medium. | High-purity NPs; no chemical precursors needed. | Expensive equipment, Low throughput, Uncontrolled aggregation, Purity issues (solvent debris) |
| Micro-emulsion | Water-in-oil microemulsion as nanoreactors for NP growth. | Highly monodisperse NPs; narrow size distribution. | Complex surfactant removal, Low concentration, Batch-to-batch variability, Toxic organic solvents |

## 2.5 The Stöber Method: Appropriate Synthesis method of Silica Nanoparticle for Electrophoretic Deposition (EPD)

The synthesis of silica nanoparticles has been explored through various methods, each with its own set of limitations. Biological synthesis, while eco-friendly, often lacks precise control over particle size and morphology [46]. Flame spray pyrolysis and laser ablation, though capable of producing high-purity nanoparticles, require specialized equipment and high temperatures, limiting their scalability and cost-effectiveness [41]. The microemulsion method, while producing highly monodisperse nanoparticles, demands large amounts of surfactants and solvents, making it less environmentally sustainable [37]. In contrast, the Stöber method (sol-gel process) overcomes many of these limitations by offering excellent control over particle size, morphology, and monodispersity through a relatively simple and scalable process. Its ability to produce uniform, spherical silica nanoparticles under mild conditions makes it particularly suitable for applications requiring precise nanomaterial properties. Furthermore, the Stöber method is highly compatible with electrophoretic deposition (EPD), as the controlled size and surface chemistry of the synthesized nanoparticles facilitate uniform deposition on substrates. This makes the Stöber method the most appropriate synthesis technique for EPD.

## 2.5.1 Stöber Method

The Stöber method is one of the most widely used techniques for synthesizing monodisperse silica nanoparticles [19]. It is a sol-gel process that involves the hydrolysis and condensation of tetraethyl orthosilicate (TEOS) in an alcohol-based solution under basic conditions. The method begins with the hydrolysis of TEOS, where water molecules react with TEOS to form silanol groups (Si-OH). These silanol groups then undergo condensation reactions to form siloxane bonds (Si-O-Si), leading to the growth of silica nanoparticles [19]. Fig 4 shows the schematic representation of silica nanoparticle synthesis by Stober method. The size and morphology of the nanoparticles can be controlled by adjusting key parameters such as the concentration of TEOS, ammonia (used as a catalyst), and the water-to-alcohol ratio [52]. The Stöber method is highly reproducible and allows for easy functionalization of the nanoparticle surface, making it suitable for applications in drug delivery, optical coatings, and catalysis [53], [54].

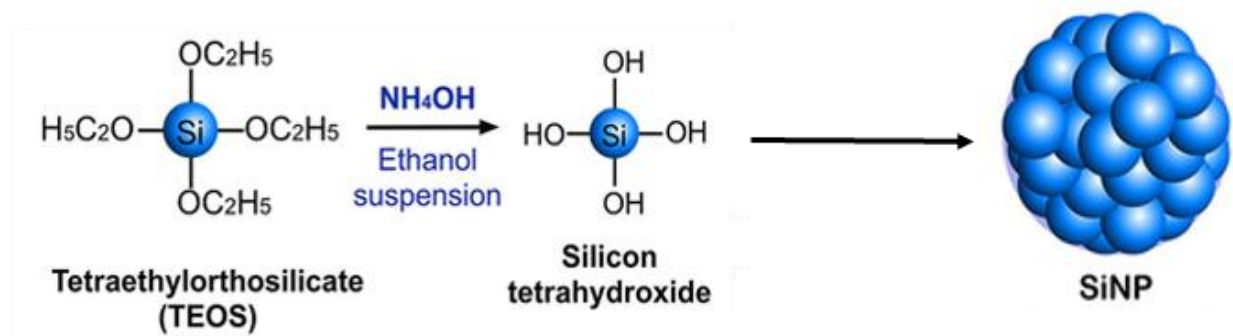

Fig 4: Schematic representation of silica nanoparticle synthesis by Stober method (Adapted from [55])

## 2.5.2 Optimal Synthesis of Silica Nanoparticle for Electrophoretic Deposition (EPD)

The Stöber method is widely regarded as one of the most effective techniques for synthesizing silica nanoparticles ($SiO_2$) for use in electrophoretic deposition (EPD) [56],[57] . This method produces highly monodisperse, spherical silica nanoparticles with tunable size and surface properties, making it ideal for EPD processes. The Stöber method's ability to control particle size, surface charge, and morphology ensures high-quality deposition when combined with EPD, which relies on the migration of charged particles under an electric field [58]. Below, we discuss why the Stöber method is particularly well-suited for silica nanoparticle deposition via EPD, supported by key facts, equations, and comparisons with other synthesis methods.

**Following are some key Advantages of Stöber Method for EPD**

a) **Monodispersity and Size Control**: The Stöber method produces silica nanoparticles with a narrow size distribution, which is critical for uniform deposition in EPD. The size of the nanoparticles can be precisely controlled by adjusting the concentration of the precursor (tetraethyl orthosilicate, TEOS), ammonia (catalyst), and the water-to-alcohol ratio. A correlation can be used to predict final particle size for concentrations 0.1-0.5 M TEOS, 0.5-17.0 M H20, and 0.5-3 M NH 3 [59]

b) **Surface Charge and Zeta Potential**: The surface charge of silica nanoparticles synthesized via the Stöber method is highly tunable, which is essential for EPD [19], [17]. The zeta potential ($\zeta$) of silica nanoparticles is typically negative due to the deprotonation of surface silanol groups (Si−OH) in aqueous solutions [60].

$$Si-OH \rightarrow Si-O^- + H^+$$

The zeta potential can be further modified by adjusting the pH or using surface functionalization agents, ensuring optimal electrophoretic mobility during EPD [61].

c) **Reproducibility and Scalability**: The Stöber method is highly reproducible, producing consistent batches of silica nanoparticles with minimal variation in size and surface properties. This reproducibility is crucial for achieving uniform deposition in EPD [62].

d) **Ease of Functionalization**: The surface of Stöber-synthesized silica nanoparticles can be easily functionalized with organic or inorganic groups (e.g., amino, thiol, or carboxyl groups) to enhance their compatibility with specific substrates or to introduce additional functionalities [63]. This is particularly advantageous for EPD, as functionalized nanoparticles can improve adhesion and film quality [64].

## 2.5.3 EPD of Stöber-Synthesized Silica Nanoparticles

The electrophoretic mobility ($\mu$) of silica nanoparticles in EPD is governed by the **Smoluchowski equation**:

$$\mu = \frac{\epsilon \zeta}{\eta}$$

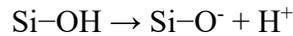

where, $\epsilon$ is the dielectric constant of the solvent, $\zeta$ is the zeta potential of the nanoparticles, $\eta$ is the viscosity of the suspension [65].

The Stöber method's ability to produce nanoparticles with a high and tunable zeta potential ensures efficient electrophoretic mobility, leading to uniform deposition. Additionally, the monodispersity of Stöber-synthesized nanoparticles minimizes agglomeration during EPD, resulting in dense, homogeneous films.

**2.6 Silica nanoparticle's multi-functionality.**

Silica nanoparticles ($SiO_2$) have emerged as a cornerstone of modern nanotechnology, celebrated for their exceptional versatility, tunable properties, and wide-ranging applications [31]. Their unique characteristics—such as high surface area, chemical stability, biocompatibility, and ease of functionalization—have positioned them as indispensable materials in advancing technologies across diverse fields [31], [66], [67]. From electronics and energy to environmental science and biomedicine, silica nanoparticles are driving innovation and addressing some of the most pressing challenges of our time. Their ability to be tailored at the nanoscale, combined with their multifunctionality, has unlocked new possibilities in material design, device fabrication, and sustainable solutions, making them a focal point of research and industrial applications. Fig 5 shows multifunctionality of Silica Nanoparticles.

In the realm of electronics, silica nanoparticles have become a key enabler of next-generation devices [68]. Their excellent dielectric properties and thermal stability make them ideal for use as insulating layers in microelectronic circuits, where they reduce leakage currents and enhance device performance [69]. Additionally, their tunable optical properties have led to their widespread use in anti-reflective coatings for displays, solar panels, and optical lenses, significantly improving light transmission and energy efficiency [70].

The field of nanofluidics has also benefited immensely from the integration of silica nanoparticles [71]. Their precise nanostructuring capabilities allow for the creation of highly controlled fluidic channels, essential for lab-on-a-chip devices used in biosensing, drug delivery, and chemical analysis [72], [46]. The porous nature of silica nanoparticles makes them excellent carriers for controlled drug release systems, where their high surface area and tunable pore size enable precise loading and sustained release of therapeutic agents [73]. Moreover, their use in nanofluidic sensors has enhanced the detection of biomolecules and environmental pollutants, offering rapid and accurate diagnostics with high sensitivity [74].

In catalysis, silica nanoparticles have proven to be invaluable as supports for catalytic materials [75]. Their high surface area and thermal stability make them ideal for heterogeneous catalysis, where they improve the activity and longevity of metal catalysts in chemical reactions [76], [77]. In photocatalysis, functionalized silica nanoparticles enhance light absorption and charge separation, enabling efficient degradation of pollutants and water splitting for hydrogen production [78]. Their role in enzyme immobilization has also opened new avenues in biocatalysis, facilitating industrial processes and environmental remediation with greater efficiency and sustainability [79].

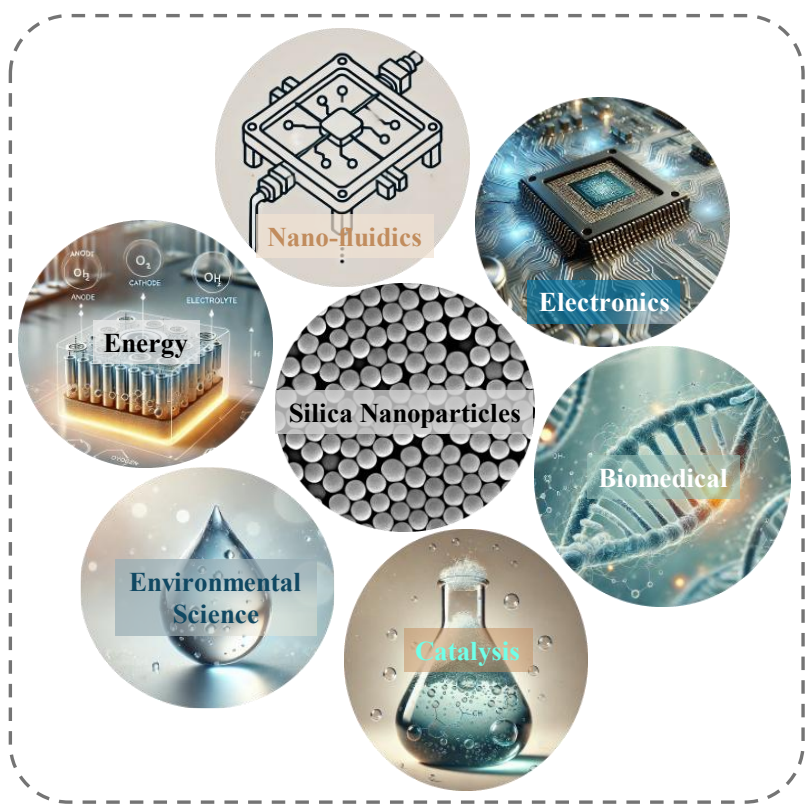

Fig 5: Multifunctionality of Silica Nanoparticles.

The energy sector has witnessed transformative advancements thanks to silica nanoparticles. In energy storage, they are used as additives in electrolytes and separators for lithium-ion batteries, improving performance and safety [80], [81]. Their high surface area and porosity have also been leveraged in supercapacitors, enhancing energy density and power output [46]. In solar energy, silica nanoparticles are incorporated into anti-reflective and light-trapping layers, boosting the efficiency of photovoltaic cells [82], [83]. Furthermore, their use as catalyst supports in fuel cells has improved stability and performance under demanding operating conditions, paving the way for cleaner and more efficient energy systems [84].

Silica nanoparticles are equally impactful in environmental science, where they address critical challenges related to pollution control and resource recovery. Their functionalized forms are used as adsorbents in water purification systems, effectively removing heavy metals, organic pollutants, and pathogens [85]. In air filtration, silica nanoparticles enhance the capture of particulate matter and volatile organic compounds, contributing to improved air quality in urban and industrial environments [86]. Their photocatalytic properties are also harnessed for the degradation of environmental pollutants, offering sustainable solutions for wastewater treatment and environmental cleanup [87], [88].

In biomedicine, silica nanoparticles have revolutionized diagnostics, drug delivery, and tissue engineering [89]. Their biocompatibility and ease of functionalization make them ideal carriers for targeted drug delivery systems, where they enable controlled release and minimize side effects [90]. In biosensing, silica nanoparticles enhance the sensitivity and specificity of diagnostic tools, allowing for early detection of diseases and pathogens [91]. Their use in tissue engineering scaffolds promotes cell growth and regeneration, while their application as contrast agents in medical imaging techniques, such as MRI and fluorescence imaging, has improved the accuracy of disease diagnosis and monitoring [92].

Beyond these fields, silica nanoparticles are making strides in coatings and surface modifications, where they enhance the durability and functionality of materials [93], [94]. Anti-corrosion coatings incorporating silica nanoparticles extend the lifespan of industrial equipment, while superhydrophobic surfaces created with these nanoparticles offer self-cleaning and water-repellent properties for textiles, construction, and automotive applications [95]. Protective coatings with silica nanoparticles provide scratch resistance, UV protection, and thermal insulation, making them invaluable in electronics, glass, and ceramics [96].

The significance of silica nanoparticles in modern technologies is highlighted by their versatility, which links fundamental research and application to real life. Their ability to be tailored for specific needs, combined with their scalability and cost-effectiveness, has made them a driving force in nanotechnology [97]. As research continues to explore new synthesis methods, deposition techniques, and application-specific designs, the potential of silica nanoparticles will only expand. Their role in addressing global challenges—from energy sustainability and environmental protection to healthcare and advanced manufacturing—highlights their transformative impact and solidifies their position as a key enabler of next-generation technologies.

**2.7. Significance of Nanoparticle Synthesis Processes on Deposition Methods**

The synthesis of nanoparticles is a critical determinant of their structural, chemical, and functional properties, which in turn influence their behavior in deposition processes and final applications. The choice of synthesis method affects particle size, shape, morphology, crystallinity, surface chemistry, and dispersibility, all of which play a pivotal role in ensuring the desired performance in nanostructured coatings, thin films, and functional nanomaterials [98].

Nanoparticles exhibit size-dependent properties, meaning even slight variations in synthesis conditions can result in substantial changes in their electronic, optical, magnetic, and catalytic behavior [99]. Parameters such as precursor concentration, reaction temperature, pH, and synthesis time determine particle size distribution, which in turn influences interparticle interactions, aggregation tendencies, and deposition uniformity [100]. Similarly, the morphology of nanoparticles, whether they are spherical, rod-shaped, or irregular, dictates their packing density and surface energy, affecting their ability to be deposited uniformly on substrates [101].

Additionally, the crystallinity and defect structures introduced during synthesis can alter the mechanical, optical, and electronic properties of nanoparticles. For instance, highly crystalline nanoparticles often demonstrate superior electrical conductivity and charge transfer properties, making them ideal for electrode materials in energy storage systems [102]. Conversely, amorphous nanoparticles may exhibit enhanced reactivity, mechanical properties and electro-catalytic performance due to the presence of more active surface sites [103].

The surface chemistry of synthesized nanoparticles is most crucial, as it dictates their stability, dispersibility, and adhesion in deposition methods [104]. Functional groups introduced during synthesis—such as hydroxyl (-OH), carboxyl (-COOH), or amine ($-NH_2$) groups—govern colloidal stability, surface charge, and interaction with deposition substrates [105]. This becomes particularly important for electrophoretic deposition (EPD), where well-dispersed nanoparticles are required to form defect-free films.

### 3. Classification of Nanoparticle Deposition Methods and their comparison

Nanoparticle deposition methods are essential for fabricating functional nanostructures in diverse applications, including sensors, electronics, catalysis, and energy devices [106]. These methods can be broadly categorized into physical, chemical, and combined approaches, each offering distinct advantages in terms of scalability, material compatibility, and deposition control [106]. In Table 2, each category is described in short, highlighting the underlying mechanisms, strengths, limitations, and applications.

### 3.1 Physical Approaches

Physical deposition methods rely on direct mechanical or physical processes to deposit nanoparticles onto substrates without altering their chemical composition. These techniques are widely employed for their high precision and uniformity, though they often face scalability, cost, and equipment complexity challenges.

**3.1.1 Sputtering:** Sputtering is a technique used to deposit thin films by ejecting material from a target (source) onto a substrate using high-energy ions (typically argon) [107]. It allows precise control over film thickness, composition, and uniformity, making it suitable for applications in semiconductors, optical coatings, and nanotechnology. Sputtering can produce dense, high-quality films with good adhesion and is compatible with a wide range of materials, including metals, oxides, and nitrides [108].

**3.1.2 Photolithography:** Photolithography is a microfabrication process used to pattern thin films by selectively exposing a photosensitive resist to ultraviolet (UV) light through a photomask. The exposed regions undergo chemical changes, allowing development to remove either the exposed or unexposed resist, creating a patterned template for etching or deposition [109]. This technique is essential in semiconductor manufacturing, MEMS, and nanodevices, enabling high-resolution patterning at the micro- and nanoscale [110].

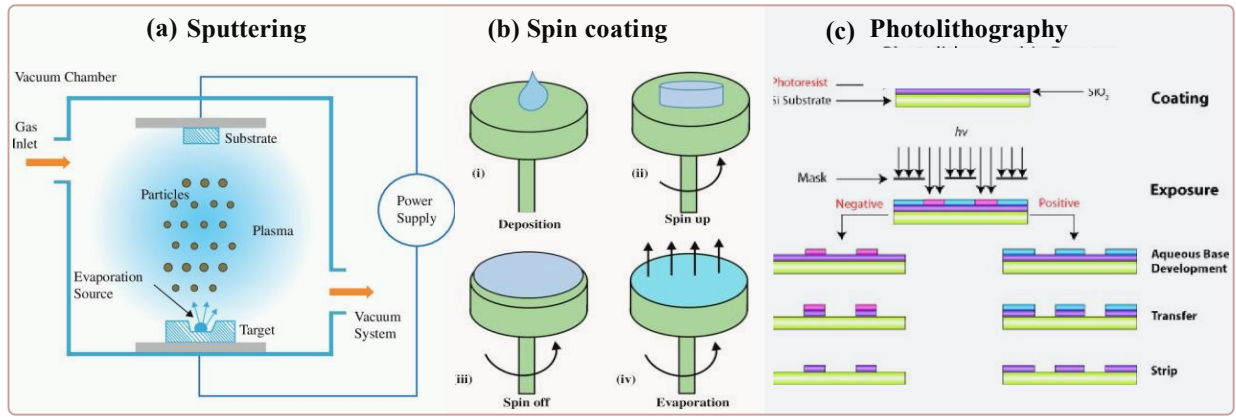

Fig 6: Schematic Diagram of different physical nanoparticle deposition processes (a) Sputtering, (b) Spin coating, and (c) Photolithography [111], [112], [113].

**3.1.3 Spin Coating and Dip Coating:** Spin coating involves dispensing a nanoparticle solution onto a rapidly rotating substrate, allowing centrifugal forces to spread and thin the material into a uniform layer [112]. Dip coating, in contrast, immerses the substrate into a nanoparticle suspension before controlled withdrawal, ensuring a uniform film [114]. While these methods are cost-effective and scalable, they suffer from poor control over nanoparticle arrangement and film thickness variations [16].

Table 2: Classification of Nanoparticle Deposition Techniques: Physical, Chemical, and Combined Methods

| Category | Technique | Process Description | Advantages | Limitations | Applications | Ref |
|---|---|---|---|---|---|---|
| Physical Approaches | Electron Beam Lithography (EBL) | Uses a focused beam of electrons to pattern nanoscale features on a substrate. | Extremely high resolution (sub-10 nm), precise control. | Slow, expensive, requires vacuum, limited to small areas. | Fabrication of nanoscale devices, photomasks, and quantum dots. | [15] |
| | Nanoimprint Lithography (NIL) | A mold with nanoscale patterns is pressed into a resist layer, which is then cured. | High throughput, high resolution, cost-effective for mass production. | Mold fabrication is expensive, limited to specific materials, alignment challenges. | Mass production of nanostructured devices, optical components, and biochips. | [115] |
| | Inkjet Printing | Droplets of ink are ejected onto a substrate to create patterns or structures. | Low cost, versatile, suitable for flexible substrates. | Limited resolution (~20 μm), ink formulation challenges, slow for large areas. | Printed electronics, biosensors, and flexible displays. | [116] |

| Technique | Description | Advantages | Disadvantages | Applications | Ref. |
|---|---|---|---|---|---|
| Dip-Pen Nanolithography (DPN) | An atomic force microscope (AFM) tip is used to deposit molecules onto a surface with nanoscale precision. | High resolution (sub-50 nm), direct writing capability. | Slow, limited to small areas, requires specialized equipment, limited scalability | Molecular electronics, biosensors, and nanoscale patterning. | [117] |
| Spin Coating | A solution is spread on a substrate by spinning, forming a thin, uniform film. | Simple, fast, good for small areas, uniform coatings, low-cost | Limited scalability, uneven thickness for large areas, material waste. | Large-area coatings, Thin-film coatings, photoresist application, and optical coatings. | [112], [16] |
| Dip Coating | A substrate is dipped into a solution and withdrawn to form a coating. | Simple, good for complex shapes, low cost. | Limited thickness control, requires post-treatment, material waste. | Anti-reflective coatings, thin films, and protective layers. | [114], [16] |
| Focused Ion Beam (FIB) | A focused beam of ions is used to mill or deposit materials at the nanoscale. | High precision, can be used for both milling and deposition, direct nanopatterning | Expensive, slow, limited to small areas, potential substrate damage. | Nanoscale machining, circuit editing, and TEM sample preparation. | [118] |
| Photolithography | Uses light to transfer a geometric pattern from a photomask to a light-sensitive chemical (photoresist) on a substrate. | High throughput, well-established, suitable for mass production. | Limited resolution (~100 nm), expensive masks, requires cleanroom facilities. | Semiconductor manufacturing, MEMS, and integrated circuits. | [109] |
| Sputtering | Energetic ions bombard a target material, ejecting atoms that deposit on a substrate. | High purity, good adhesion, versatile for various materials. | High cost, requires vacuum, limited scalability. | Thin-film deposition, optical coatings, and semiconductor devices. | [107], [108] |
| Thermal Evaporation | Material is heated to vaporize and condense on a substrate. | Simple vapor deposition, high purity, good for thin films. | Limited to materials with low melting points, poor step coverage, requires vacuum. | Thin-film deposition, metallization, and organic electronics. | [119] |
| Soft Lithography | Uses elastomeric stamps or molds to pattern materials on a substrate. | Low cost, suitable for flexible substrates, high resolution. | Limited to specific materials, stamp wear and tear, alignment challenges. | Microfluidics, biosensors, and flexible electronics. | [120] |
| Self-Assembly | Molecules or nanoparticles spontaneously organize into ordered structures. | Simple, cost-effective, Large-scale, defect-free nanoparticle organization | Limited control over final structure, requires specific molecular interactions. | Nanostructured materials, photonic crystals, and drug delivery systems. | [121] |

| | | | | | | |
|---|---|---|---|---|---|---|
| **Chemical Approaches** | Physical Vapor Deposition (PVD) | Material is vaporized from a solid source and deposited onto a substrate. | High purity, good adhesion, versatile. | Requires vacuum, limited scalability, high cost, slow. | Thin-film coatings, wear-resistant coatings, and optical devices. | [10] |
| | Chemical Vapor Deposition (CVD) | Gaseous precursors react on a substrate to form a solid deposit. | Conformal coatings, high-quality films, versatile. | High temperatures, complex setup, expensive precursors. | Thin-film deposition, semiconductor devices, and protective coatings. | [11] |
| | Atomic Layer Deposition (ALD) | Sequential, self-limiting reactions form atomic layers on a substrate. | Atomic-level precision, excellent uniformity, conformal coatings. | Slow process, expensive equipment, limited material choices, limited to thin films | High-k dielectrics, MEMS, and nanolaminates, Semiconductor devices, ultra-thin films | [14] |
| | Sol-Gel Process | Precursor solution undergoes hydrolysis and condensation to form a gel, which is then dried to create a solid material. | Low cost, versatile, good for large areas. | Requires post-treatment, limited thickness control, long processing times. | Thin films, coatings, and porous materials. | [122] |
| | Electrospinning | A high-voltage electric field is used to draw fibers from a polymer solution or melt. | Produces nanofibers, versatile, scalable. | Limited to polymers, requires optimization of solution properties. | Tissue engineering, filtration, and drug delivery, energy storage | [123] |
| | Langmuir-Blodgett (LB) Deposition | Monolayers of molecules are transferred from a liquid surface to a solid substrate. | High precision, uniform monolayers, good for organic materials. | Slow, limited to specific materials, requires specialized equipment. | Organic electronics, biosensors, and thin-film devices. | [124] |
| **Combined Approaches** | Electrochemical Deposition | Uses an electric current to reduce metal ions from a solution onto a conductive substrate. | Low cost, scalable, good for metals and alloys. | Limited to conductive substrates, requires precise control of parameters. | Metal coatings, batteries, and sensors, energy storage devices | [125] |
| | Directed Self-Assembly (DSA) | Combines self-assembly with external guidance (e.g., templates or fields) to create ordered nanostructures. | High resolution, scalable, cost-effective. | Requires precise control of conditions, limited to specific materials. | Nanoscale patterning, photonics, and semiconductor devices. | [126] |
| | Nanoimprint Self-Assembly | Combines nanoimprint lithography with self-assembly to create nanostructures. | High resolution, scalable, cost-effective. | Limited to specific materials, requires precise control of conditions. | Nanostructured materials, photonic devices, and sensors. | [127] |

| | Electrohydrodynamic Jet printing (e-jet) | Uses an electric field to eject droplets from a nozzle, enabling high-resolution printing. | High resolution (sub-100 nm), versatile. | Requires precise control of parameters, limited to specific inks. | Printed electronics, biosensors, and nanoscale patterning. | [128] |
|---|---|---|---|---|---|---|
| | Plasma-Enhanced CVD (PECVD) | Uses plasma to enhance chemical reactions in CVD, enabling lower-temperature deposition. | Lower temperatures, faster deposition, conformal coatings. | Expensive, requires specialized equipment, limited material choices. | Thin-film transistors, solar cells, and protective coatings. | [129] |

## 3.2 Chemical Approaches

Chemical deposition techniques utilize solution-based or gas-phase reactions to deposit nanoparticles onto substrates, offering advantages in scalability and cost-effectiveness. However, these methods may require post-treatment to achieve uniformity and desired material properties.

**3.2.1 Physical Vapor Deposition (PVD):** PVD encompasses techniques such as sputtering and thermal evaporation, where nanoparticles or thin films are deposited through the vaporization of a target material in a vacuum environment [10]. This method enables uniform coatings with excellent adhesion, making them suitable for optics, microelectronics, and protective coatings applications [130]. However, PVD typically requires high-vacuum conditions and additional masking for controlled deposition, increasing complexity and operational costs [10].

**3.2.2 Chemical Vapor Deposition (CVD):** CVD is a widely used technique in which gaseous precursors react at elevated temperatures to form thin films of nanoparticles on a substrate. This method enables the fabrication of highly conformal coatings with excellent compositional control [11]. However, the high temperatures and complex processing requirements can limit its applicability, particularly for temperature-sensitive materials [131].

**3.2.3 Atomic Layer Deposition (ALD):** Atomic Layer Deposition (ALD) is a vapor-phase chemical deposition technique that enables precise atomic-scale control over thin-film growth [14]. It relies on sequential, self-limiting surface reactions, where precursor gases alternately react with the substrate in a layer-by-layer fashion, ensuring uniform, conformal coatings. ALD is widely used in semiconductors and nanotechnology for high-precision coatings, though its slow deposition rate and high-cost limit large-scale applications [132].

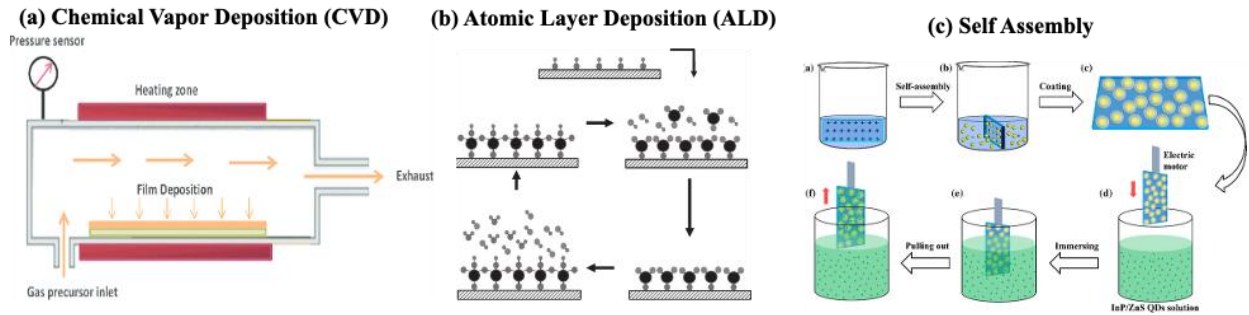

Fig 7: Schematic Diagram of different Chemical nanoparticle deposition processes [133], [132], [134]

**3.2.4 Self-Assembly Techniques:** Self-assembly methods leverage molecular interactions such as electrostatic, van der Waals, or hydrophobic forces to organize nanoparticles into structured films. While this approach enables large-scale deposition with minimal equipment costs, controlling the precise arrangement of nanoparticles remains a challenge [121].

**3.2.5 Electrospinning:** Electrospinning employs electrostatic forces to generate nanofibers embedded with nanoparticles from a polymer solution. This technique produces highly porous and high-surface-area coatings beneficial for energy storage and filtration applications. However, achieving uniform nanoparticle distribution within the fibers requires careful optimization of solution properties and processing parameters [123].

**3.3 Combined Approaches**

Hybrid deposition methods integrate physical and chemical techniques to leverage their respective advantages, enabling greater control over nanoparticle deposition and film characteristics.

**3.3.1 Plasma-Enhanced Chemical Vapor Deposition (PECVD):** PECVD introduces plasma energy to conventional CVD processes, enabling lower-temperature deposition while improving film adhesion and uniformity. This technique is beneficial for depositing functional coatings on temperature-sensitive substrates. Nevertheless, its reliance on specialized equipment increases operational costs [129].

**3.3.2 Nanoimprint Self-Assembly:** Recent advances have integrated nanoimprint lithography (NIL) with self-assembly techniques to enhance nanofabrication resolution and functionality. In hybrid patterning, NIL creates master templates while self-assembly refines finer sub-patterns, enabling sub-10 nm features. Directed self-assembly (DSA) further improves precision by using NIL-defined chemical guides to align block copolymers or colloidal nanoparticles [127]. However, challenges remain, including defect control in large-area patterning, material compatibility limitations, and the need for stringent process optimization to ensure reproducibility at industrial scales [127].

Combined approaches effectively balance the advantages of physical and chemical methods, offering precise control over nanoparticle deposition while maintaining scalability and cost-effectiveness. These methods are particularly promising for next-generation nanofabrication applications, where tailored coatings and functional interfaces are crucial [135].

## 4. Electrophoretic Deposition (EPD) for Fabricating Nanofluidic Devices and Nanoporous Structures with Silica Nanoparticles

Electrophoretic deposition (EPD) has garnered significant attention in nanofabrication due to its superior capability to deposit silica nanoparticles with high precision and efficiency [17]. Compared to conventional deposition techniques, EPD offers uniform deposition, precise porosity control, room-temperature operation, scalability, and cost-effectiveness [17]. This section delves into the fundamental principles of EPD, its advantages over competing methods, and its broad applicability in nanofluidic and nanoporous structure fabrication.

**4.1. Fundamentals Principles of Electrophoretic Deposition (EPD):** Electrophoretic Deposition (EPD) is a versatile and efficient technique for depositing nanoparticles, including silica nanoparticles, onto conductive substrates. The process relies on the movement of charged particles suspended in a colloidal solution under the influence of an applied electric field. When an electric field is applied, the charged particles migrate toward the oppositely charged electrode (substrate), where they deposit and form a dense, uniform coating [136]. EPD is governed by two main processes: electrophoresis (movement of charged particles) and deposition (particle accumulation on the substrate) [17]. The quality of the deposited film depends on factors such as particle charge, suspension stability, electric field strength, and deposition time. Fig 8 shows a schematic of the Electrophoretic Deposition Process (EPD) for silica nanoparticle deposition.

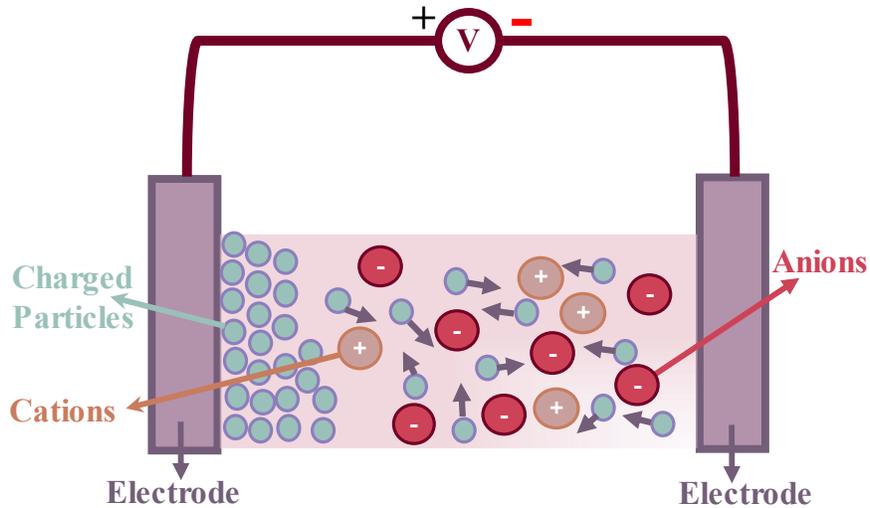

Fig 8: Schematic of Electrophoretic Deposition Process (EPD) for silica nanoparticle deposition [137].

### 4.2. Parameters Affecting EPD of Silica Nanoparticles

Several key parameters influence the final deposition of silica nanoparticles during EPD. Fig shows how different parameters affect the end results of EPD deposition.

**4.2.1 suspension parameters:** The EPD process is governed by critical suspension characteristics: zeta potential (±30 mV for stability), conductivity (balancing migration and electrolysis), particle size/distribution (affecting mobility and packing density), and dielectric constant (influencing double-layer polarization) [138], [139]. Optimal control of these parameters ensures uniform deposition, while improper tuning may lead to defects, agglomeration, or non-uniform coatings. A systematic understanding of their interplay is essential for reproducible fabrication of high-performance nanostructured films [139].

**4.2.2. Process Parameter:** The EPD process is critically controlled by several operational parameters like applied voltage (typically 1-100 V) determines particle migration rate and affects film density, with excessive voltage potentially causing bubble formation; deposition time governs film thickness, though prolonged durations may lead to uneven growth or cracking; substrate conductivity influences deposition uniformity, with conductive surfaces enabling homogeneous coatings while insulating substrates require modified approaches; and nanoparticle concentration (usually 0.1-10 g/L) directly impacts deposition rate and final film morphology [17].

Optimal parameter selection is essential for achieving tailored nanostructured coatings with desired properties, requiring careful balance between deposition kinetics and film quality [17].

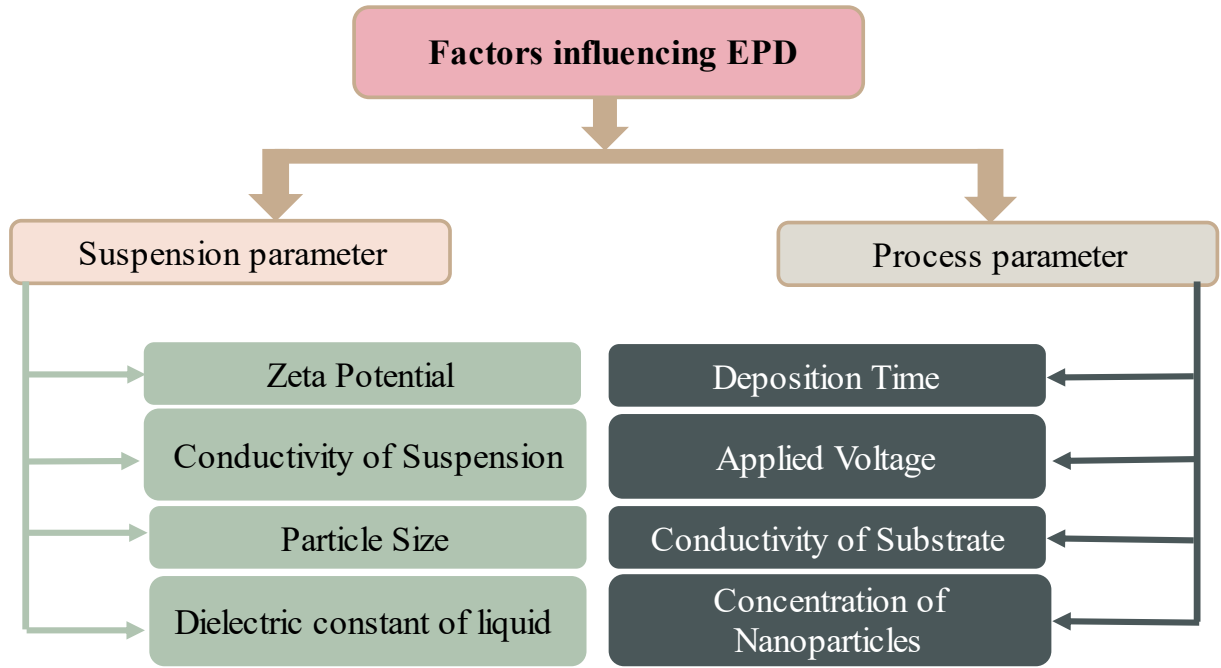

Fig 9: Factors influencing EPD [17].

## 4.3. Advantages of Electrophoretic Deposition (EPD) Over Conventional Deposition Techniques

Electrophoretic deposition (EPD) stands out as an excellent technique for fabricating nanofluidic devices and nanoporous structures due to its unique combination of uniformity, scalability, versatility, and environmental sustainability. Below, Fig 10 shows a comparative analysis highlighting EPD's advantages over conventional physical, chemical, and hybrid deposition methods

EPD enables homogeneous coatings with controlled pore size distribution, which is critical for nanofluidic applications requiring precise fluid transport. Unlike spin coating and dip coating which suffer from uneven aggregation or self-assembly lacking structural control, EPD ensures smooth, continuous nanoporous layers [16], [112], [114], [136]. EPD operates at ambient temperatures, unlike high-temperature methods chemical vapor deposition (CVD), Atomic layer deposition (ALD), and thermal evaporation that degrades heat-sensitive substrates like polymers, biomaterials, etc. [11], [14].

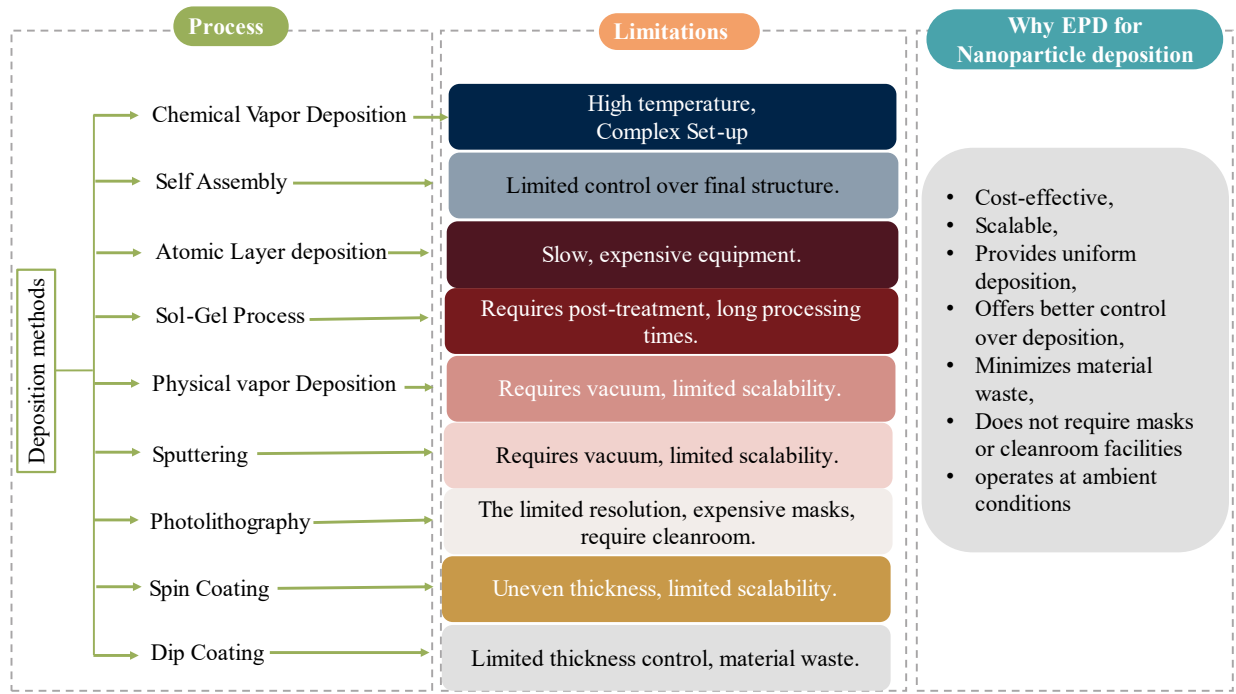

Fig 10: Comparison of EPD vs major common conventional deposition methods [11], [121], [14], [122], [10], [107], [108], [109], [112], [16], [114].

EPD also preserves nanoparticle properties and substrate integrity, making it ideal for biomedical and flexible electronics applications [140]. It does not require vacuum systems or complex infrastructure, unlike Electron Beam Lithography (EBL), Focused Ion Beam (FIB), or Physical Vapor Deposition (PVD) [10], [15], [118]. Also it does not need costly molds like in Nanoimprint Self-Assembly and cleanroom dependencies like for photolithography [109], [127]. Large-area deposition capability outperforms dip-pen nanolithography (DPN) and inkjet printing which is limited by resolution and speed [116], [117]. EPD offers exceptional substrate versatility and material compatibility, enabling uniform deposition on planar, curved, and 3D substrates, unlike nanoimprint lithography (NIL) or sputtering [107],[108],[115]. Also Compatible with functionalized silica nanoparticles for tailored surface chemistry, a limitation in PVD and sol-gel processes [10], [60], [122]. EPD is an environmentally sustainable deposition method that utilizes aqueous or organic solvents, minimizing hazardous byproducts and material waste compared to conventional chemical vapor deposition (CVD), and Plasma-Enhanced CVD

(PECVD) [10], [11], [129]. EPD also minimizes material waste compared to spin coating or inkjet printing [16], [116]. EPD's cost-effectiveness, ambient processing, and adaptability make it the preferred choice for industrial-scale nanofabrication, particularly for silica nanoparticle deposition. By addressing the shortcomings of conventional methods—such as non-uniformity (spin coating), high-temperature constraints (CVD), and scalability limits (EBL) EPD bridges the gap between lab-scale research and commercial applications [139].

## 5. Future Work

**AI-Driven Optimization of Electrophoretic Deposition (EPD) for Silica Nanoparticles**

The Electrophoretic Deposition (EPD) process has emerged as a versatile and scalable technique for depositing silica nanoparticles ($SiO_2$) onto various substrates, enabling applications in energy storage, biomedicine, catalysis, and environmental science [141], [17]. However, optimizing EPD parameters—such as voltage, deposition time, suspension pH, and particle concentration—remains a complex and time-consuming task, often requiring extensive trial-and-error experimentation. Recent advancements in Artificial Intelligence (AI) with Active Learning offer transformative opportunities to streamline this optimization process, enhance deposition precision, and unlock new possibilities for advanced material design [142], [143], [144]. This section explores how AI and active learning can revolutionize EPD, providing a roadmap for future research and industrial applications [145]. Fig 11 shows how AI-driven optimization can help to improve the EPD of non-metallic nanoparticles like silica nanoparticles.

**5.1 Development of Comprehensive Datasets for AI Training**

The successful integration of artificial intelligence (AI) in electrophoretic deposition (EPD) optimization relies on high-quality, comprehensive datasets that capture the intricate relationships between process parameters, suspension properties, and film characteristics [146]. Future research must prioritize standardized data collection, open-access repositories, and high-throughput experimental techniques to enhance AI model training [147]. Emphasis should be placed on surface chemistry data (e.g., zeta potential, pH, ionic density) and suspension parameters (e.g., electrolyte conductivity, particle size), as these critically influence deposition behavior and film quality [148]. Additionally, multiple optimization objectives must be considered to ensure films meet specific application requirements, such as deposition uniformity, surface roughness, porosity, and adhesion strength [17]. Standardized data collection and open-access repositories are essential to build robust datasets for AI-driven electrophoretic deposition (EPD) optimization [149]. Developing standardized protocols for measuring key parameters such as zeta potential, pH, electrolyte conductivity, and film properties ensures data consistency and reproducibility [150]. Creating open-access databases facilitates data sharing, fostering collaboration and improving the generalizability of AI models [151]. Additionally, high-throughput experimentation can enrich data collection by automating deposition processes, enabling multiple experiments under varying conditions, and incorporating real-time monitoring tools to track critical parameters such as zeta potential, conductivity, and film properties like thickness and roughness [152]. Integrating experimental data with computational models further enhances dataset comprehensiveness, enabling more accurate AI training [153].

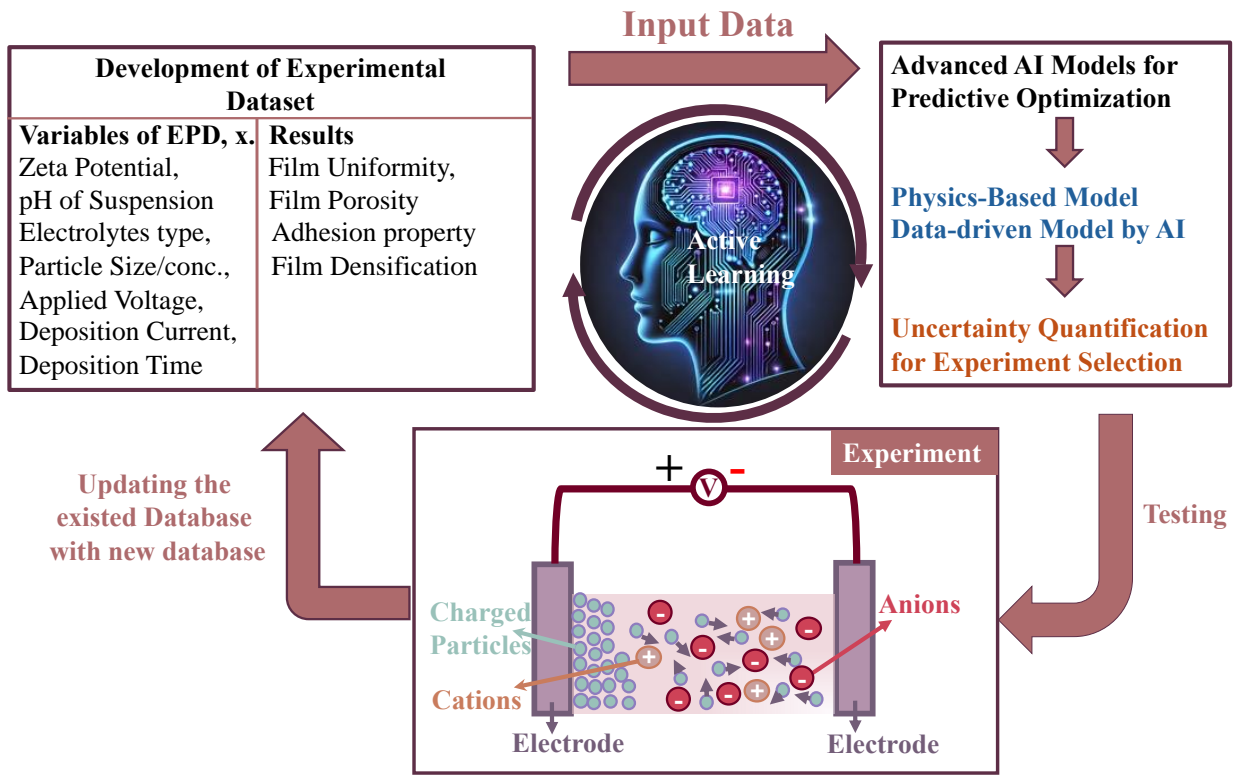

Fig 11: AI-Driven Optimization Workflow for Electrophoretic Deposition (EPD) (adapted from [154], [155, p.], [154] ).

## 5.2 Advanced AI Models Development for Predictive Optimization

Developing and refining advanced AI models are crucial for predicting optimal electrophoretic deposition (EPD) conditions and film properties with desirable accuracy [156]. Future research should focus on integrating hybrid modeling approaches, transfer learning techniques, and uncertainty quantification to enhance predictive capabilities and experimental efficiency.

A synergistic combination of physics-based models such as Electrochemical Impedance Spectroscopy (EIS) Models, Finite Element Method (FEM) for Electrodeposition and data-driven machine-learning techniques like Gaussian Process Regression (GPR), Neural Networks (NNs), Bayesian Networks can improve the interpretability and reliability of AI predictions [145]. By incorporating fundamental electrochemical and colloidal interactions into AI frameworks, these hybrid models can better capture the complex dependencies between deposition parameters and film characteristics, reducing reliance on extensive experimental datasets [157]. Leveraging pre-trained models from related domains such as thin-film deposition, colloidal science, and fluid dynamics can significantly accelerate the development of AI-driven EPD optimization tools [158]. Incorporating uncertainty quantification into AI models can identify regions within the parameter space where predictions exhibit higher uncertainty, guiding the selection of the most informative experiments [159].

## 5.3 Active Learning for Efficient Experimentation

Active learning offers a powerful approach to optimizing EPD by iteratively selecting the most informative experiments [160], [161]. Future work should explore Adaptive Sampling Strategies, Multi-Objective Optimization, & Integration with Robotics. Developing adaptive sampling algorithms that dynamically adjust experimental plans based on real-time data and model predictions [162]. Extending active learning to multi-objective optimization, where multiple film properties like thickness, uniformity, and adhesion are optimized simultaneously [163]. Combining active learning with robotic experimental platforms to automate the selection and execution of experiments, further reducing costs and time. These efforts will enhance the efficiency of EPD optimization, enabling faster discovery of optimal deposition conditions.

**5.4 Real-Time Optimization and Future Applications**

Real-time optimization in electrophoretic deposition (EPD) leverages AI-driven models to dynamically adjust process parameters (e.g., voltage, pH, nanoparticle concentration) based on in-situ data, enhancing deposition uniformity and efficiency. Machine learning algorithms, such as reinforcement learning, can autonomously refine EPD conditions to achieve optimal film properties. Future applications include scalable nanomanufacturing of advanced coatings, energy storage materials, and biomedical devices, where AI-enabled EPD could enable precise control over nanostructured surfaces.

To realize this potential, challenges in scalability and industrial adoption must be addressed. Key focus areas include Process Scaling, developing EPD systems adaptable to large-area coatings and high-volume production; Cost Reduction, optimizing cost-effectiveness by minimizing reliance on expensive equipment or materials; and Industry Collaboration, partnering with manufacturers to validate AI-driven EPD in real-world applications for example in electronics, energy, healthcare. By bridging the gap between lab-scale research and industrial needs, these efforts will pave the way for widespread adoption of AI-optimized EPD in next-generation nanofabrication.

The integration of AI and active learning into the EPD process represents a transformative opportunity to advance nanotechnology. By addressing challenges related to data quality, model development, real-time control, material design, scalability, and ethical considerations, future research can unlock the full potential of AI-driven EPD. These efforts will enable the development of high-performance materials with tailored properties, paving the way for innovative applications in energy storage, biomedicine, environmental science, and beyond. As AI-driven EPD continues to evolve, it will play a pivotal role in shaping the future of nanotechnology that benefits society as a whole.

## 6. Conclusion

This review paper provides a perspective on nanoparticle deposition techniques, particularly for silica nanoparticles ($SiO_2$), which have emerged as pivotal tools in advancing nanoscale architectures for applications spanning electronics, energy storage, biomedicine, and environmental science. This review critically examined the synthesis of $SiO_2$ nanoparticles via the Stöber method, which offers unique control over particle size, mono-dispersity, and surface

charge—key attributes that make it appropriate for electrophoretic deposition (EPD). EPD stands out among deposition techniques due to its scalability, cost-effectiveness, room-temperature processing, and ability to produce uniform, high-quality films on complex substrates.

The transformative perspective for AI-driven optimization in EPD in further research is a major feature of this study. Researchers can overcome traditional trial-and-error limitations by leveraging machine learning, active learning, and real-time adaptive control, enabling predictive modeling, and dynamic parameter adjustment. Future advancements in high-throughput experimentation, hybrid physics-AI models, and industrial collaboration will be crucial in transitioning AI-optimized EPD from lab-scale research to large-scale manufacturing. Despite these advancements, challenges remain in process scalability, cost reduction, and material compatibility, particularly for non-metallic nanoparticles. Addressing these hurdles through sustainable synthesis methods, standardized datasets, and ethical AI integration will be essential for widespread adoption. The convergence of Stöber-synthesized $SiO_2$, EPD, and AI-driven optimization in nanofabrication offers unique precision and efficiency in designing next-generation functional materials. As the field progresses, the synergy between nanoparticle engineering, advanced deposition techniques, and intelligent automation will unlock innovative solutions for global challenges in energy, healthcare, and environmental sustainability. This review consolidates current knowledge and charts a roadmap for future research, emphasizing the need for interdisciplinary collaboration to harness the full potential of AI-enhanced nanomanufacturing.